\documentclass[]{revtex4}
\usepackage{epsf,epsfig}

\begin{document}

\title{Functional Bias and Spatial Organization of Genes in Mutational
  Hot and Cold Regions in the Human Genome} \author{Jeffrey H.
  Chuang$^1$} \author{Hao Li$^1$} \affiliation{$^1$Department of
  Biochemistry and Biophysics, University of California, San
  Francisco, California 94107 USA}

\begin{abstract}
  {\bf Background:} The neutral mutation rate is known to vary widely
  along human chromosomes, leading to mutational hot and cold regions.
  
  {\bf Methodology/Principle Findings:} We provide evidence that
  categories of functionally-related genes reside preferentially in
  mutationally hot or cold regions, the size of which we have
  measured. Genes in hot regions are biased toward extra-cellular
  communication (surface receptors, cell adhesion, immune response,
  etc.) while those in cold regions are biased toward essential
  cellular processes (gene regulation, RNA processing, protein
  modification, etc.).  From a selective perspective, this
  organization of genes could minimize the mutational load on genes
  that need to be conserved and allow fast evolution for genes that
  must frequently adapt. We also analyze the effect of gene
  duplication and chromosomal recombination, which contribute
  significantly to these biases for certain categories of hot genes.
  
  {\bf Conclusions/Significance:} Overall, our results show that genes
  are located non-randomly with respect to hot and cold regions,
  offering the possibility that selection acts at the level of gene
  location in the human genome.
\end{abstract}

\maketitle

\section{Introduction} 

Because of the abundant availability of mouse and human genome data
(Mouse Genome Sequencing Consortium 2002, International Human Genome
Sequencing Consortium 2001), it has come to light that mutation rates
vary widely across different regions of the human genome (Hardison et
al. 2003, Mouse Genome Sequencing Consortium 2002, Matassi et al.
1999), in agreement with a number of smaller-scale studies (Wolfe et
al. 1989, Perry and Ashworth 1999, Casane et al.  1997).  Regions of
unusually high or low substitution rates have been observed from
4-fold sites and ancestral repeat sequences, two of the best
candidates for measuring neutral rates of mutation in mammals
(Hardison et al. 2003, Mouse Genome Sequencing Consortium 2002, Sharp
et al.  1995).  The reasons for such regional variability are unclear,
since structural characterizations of the mutation rate are nascent.
Whatever the reason for these hot and cold regions, their existence
suggests a question that has intriguing consequences for molecular
evolution: does the organism take advantage of these hot and cold
spots?

One way to take advantage of a hot region would be to place genes
there for which the hotness is useful -- an intuitive example would be
receptor proteins, which must respond to a constantly changing ligand
set. Similarly, it could be beneficial to place delicate genes in a
cold region, to reduce the possibility of deleterious mutations. These
potential advantages offer the possibility that regional mutation
rates affect the spatial organization of genes.  The idea of such
organization in mouse and human is bolstered by recent findings of
gene organization in yeast. For example, Pal and Hurst showed that
yeast genes are organized to take advantage of local recombination
rates (Pal and Hurst 2003), which is particularly relevant since
mutation rate and recombination rate are known to be correlated
(Lercher and Hurst 2002).  If the local mutation rate -- equivalent to
the synonymous (amino-acid preserving) substitution rate $K_S$ if
synonymous substitutions are neutral -- affects gene organization,
this would constitute a type of selection complementary to traditional
selection on point mutations (Graur and Li 2000) .

We studied whether local mutation rates affect gene locations by
measuring the mutation rates of genes and their organization in the
human genome. First, we analyzed the substitution
rates of the genes in each of the families defined by the Gene Ontology 
(Gene Ontology Consortium 2000) .
If the organism is taking advantage of varying $K_S$, gene families
should be biased toward regions of appropriate rate. In fact we
observe that several functional classes of genes preferentially occur
in hot or cold regions.  Some of the notable hot categories we observe
are olfactory genes, cell adhesion genes, and immune response genes,
while the cold categories are biased toward regulatory proteins such
as those involved in transcription regulation, DNA/RNA binding, and
protein modification. Also, to better characterize the hot and cold
regions, we measured the length scale over which substitution rates
vary. While rough bounds on the size of hot and cold regions are known
(Hardison et al.  2003, Matassi et al. 1999), this paper presents the
first quantitative calculation of their length scale.

Because mutation rates are regional, mutation rates in genes
categories could be influenced by events altering the organization of
genes in the genome, such as gene relocation or gene duplication.  We
therefore analyzed mechanisms by which functional categories of genes
may have become concentrated in hot or cold regions.  A clustering
analysis reveals that the hotness of some categories is enhanced by
local gene duplications in hot regions.  However, there are strong
functional similarities among the hot categories -- both clustered and
unclustered, as well as among the cold categories.  These functional
similarities imply that the instances of duplicated categories are not
random. i.e.  selection may have affected which genes have duplicated
and persisted.

\section{Results}

\subsection{Mutation Rates Have Regional Biases}

Recently, substitution rates between {\em Mus musculus} and {\em Homo
  sapiens} have been measured by several groups on a genome-wide scale
(Mouse Genome Sequencing Consortium 2002, Hardison et al. 2003, Kumar
and Subramanian 2002).  These substitution rates vary significantly
across the genome (Mouse Genome Sequencing Consortium 2002, Hardison
et al.  2003), suggesting that neutral mutation rates may have
regional biases as well. A popular proxy for neutral mutation rates is
the substitution rate at 4-fold sites (a recent example is (Kumar and
Subramanian 2002)), base positions in coding DNA which do not affect
protein sequence, and which should hence be under less selective
pressure than other sites. The 4-fold sites also offer the advantage
of being easily alignable.

For these reasons, we estimated the neutral mutation rate from
substitution rates at 4-fold sites (which we use interchangeably with
the term $K_S$ in this paper). This identification is not without
complexities, however, since there are processes which can in
principle selectively affect the 4-fold sites. For example, some have
argued that exogenous factors such as isochore structure influence the
silent sites (Bernardi 2000), and codon usage adaptation has been
shown to affect silent sites in bacteria and yeast (Sharp and Li 1987,
Percudani and Ottonello 1999). So far, such selective effects have
been difficult to detect in mammals (Iida and Akashi 2000, Kanaya et
al. 2001, Duret and Mouchiroud 2000, Smith and Hurst 1999a).
Recently, Hardison et al showed that several functionally unrelated
measures of mutation rate, including SNP density, substitutions in
ancestral repeats, and substitutions in 4fold sites, are correlated in
genome-wide mouse-human comparisons (Hardison et al. 2003) --
suggesting that these measures have common neutral aspects.

We constructed our own dataset of the 4-fold substitution rates for
14790 mouse/human orthologous genes, using data from the Ensembl
consortium. In order to properly account for stochastic finite-size
effects, we mapped the observed substitution rates to a normalized
value, based on the assumption that all 4-fold sites mutate at the
same rate (see Methods).  Under this assumption, it was expected that
the normalized substitution rates would follow the Normal distribution
(a gaussian with $\sigma = 1$).

Contrary to these expectations, the distribution of ortholog
substitution rates was found to be highly biased toward high or low
rates, indicating that 4-fold mutation rates vary substantially by
location, and on a scale larger than the typical size of a gene.  Fig.
\ref{fig:mutdistribution} shows the distribution of substitution rates
for all mouse/human orthologs.  The observed distribution has excesses
of genes at both high and low substitution rates. These results are in
agreement with the findings of Matassi et al (Matassi et al. 1999),
who reported significant mutation rate correlations between
neighboring genes.  This is not a compositional effect -- the
distribution remained the same even when corrections for the gene's
human base composition were made (see Methods).  We further verified
that substitution rates of neighboring genes were correlated using an
analysis qualitatively similar to Matassi et al -- though with
approximately 20 times more orthologs -- finding that gene
substitution rates are correlated with their neighbors with a p-value
of $10^{-189}$ (See Methods).  These results imply that substitution
rates have regional biases acting both within a gene and over longer
length scales.

\subsection{Some Gene Categories Are Biased Toward Hot or Cold Regions}

We next considered whether there is a relationship between gene
locations and their functions.  i.e. whether functional categories of
genes have biases for being in regions of particular mutation rate.
To test whether such biases exist, we performed an analysis of the
Gene Ontology (GO) assignments for each ortholog pair (Gene Ontology
Consortium 2000), using data from the Ensembl human ENSMART database
to assign genes to GO categories.
For each GO category, we calculated a $z$-score to measure the overall
substitution rate, based on the substitution rates of the genes in the
category (see Methods).  The 21 GO categories having statistically
significant positive values of $z$ are shown in Table
\ref{table:GOhighsub}.  In terms of 4-fold substitution rates, the hot
category rate averages were found to range from 0.346 (integral to
membrane) to 0.468 (internalization receptor activity), while the
genome-wide average was 0.337 (with gene-wise standard deviation
0.08). For a category with several genes, the effective standard
deviation is much smaller, equal to $0.08/\sqrt{N_{GO}}$, where
$N_{GO}$ is the number of genes in the category; so these rate biases
are extremely significant.  Hot gene categories were focused mainly in
receptor-type functions, along with a few other categories such as
proteolysis and microtubule motor activity. Some preferences were
partially because categories have genes in common, e.g. 8 genes are
shared between the categories dynein ATPase activity, dynein complex
and microtubule-based movement category.  However, there were several
categories of similar function which were independent. e.g.  membrane
and olfactory receptor activity shared no genes, and cell adhesion and
immune response shared only 5\% of their genes. Overall, there was a
clear bias for the larger hot categories to contain receptor-type
proteins: e.g.  receptor activity, olfactory receptor activity,
G-protein coupled receptor protein signaling pathway, membrane, and
immune response.  For the set of all genes where the string
``receptor'' is part of the GO description, the average 4-fold
substitution rate was found to be 0.347. The probability for the 
set of 1488 receptor genes to have a mutation rate this high is
$10^{-6}$.

The 36 statistically significant GO categories with negative $z$
scores, are shown in Table \ref{table:GOlowsub}.  The 4-fold rate
averages for the cold categories ranged from 0.220 (mRNA binding
activity) to 0.326 (protein serine/threonine kinase activity).  The
coldest gene categories included nuclear proteins, transcription
regulation, DNA and RNA binding, oncogenesis, phosphatases, and
kinases, all of which are important to regulatory processes. Many of
these genes are also housekeeping genes (Hsiao et al. 2001).  For the
set of all genes where the string ``regulat'' is part of the GO
description, the average 4-fold substitution rate was found to be
0.325. The probability for the 
set of 1704 regulation genes to have a mutation rate this high is
$10^{-9}$.

We repeated our $z$-score classifications using several other measures
of mutation rate and in each case inferred similar hot and cold
categories.  For example, under the normalized rate model that
accounts for human base composition, the same set of 23 hot categories
were found. Of the 37 cold categories, 33 remained classified as cold.
The 4 lost were: regulation of transcription from Pol II promoter,
development, neurogenesis, and translation regulator activity. There
were 6 new categories, and these were also largely regulatory: nucleic
acid binding activity, translation initiation factor activity,
ubiquitin C-terminal hydrolase activity, collagen, RNA processing, and
negative regulation of transcription.  We also calculated several
maximum likelihood (ML) measures of $K_S$ using mutation models in the
PAML package (Yang 1997), including the Nei and Gojobori (Nei and
Gojobori 1986) codon-based measure and the Tamura-Nei (TN93) and REV
(Tavere 1986) models. We again found qualitatively similar sets of hot
and cold categories -- receptor genes at high substitution rates and
regulatory genes at low substitution rates -- though there were
changes in the numbers of significant categories.  For example, for
the TN93 model, we observed 10 hot categories: induction of apoptosis
by extracellular signals, G-protein coupled receptor protein signaling
pathway, olfactory receptor activity, receptor activity, apoptosis,
enzyme activity, chymotrypsin activity, trypsin activity, integral to
membrane, and dynein ATPase activity; and 8 cold categories:
calcium-dependent protein serine/threonine phosphatase activity,
ribonucleoprotein complex, protein serine/threonine kinase activity,
RNA binding activity, protein amino acid dephosphorylation,
intracellular protein transport, protein transporter activity, and
nucleus. The categories inferred from our original $z$-score analysis 
are probably  more accurate than those from ML methods, because ML methods 
tend to produce strong outliers at high substitution rate, skewing
calculations of the variance in the $z$-score analysis.

\subsection{Can Gene Duplications Explain the Hot and Cold Categories?}

Given the existence of hot and cold gene categories, the question then
becomes: why do these biases exist?  One potentially non-selective
factor that could affect category rate biases is local gene
duplications.  New genes generally arise by duplication, in which a
new copy of a gene is generated nearby to the pre-existing gene by a
recombinatorial event such as unequal crossing-over, followed by
evolution to a novel, but often related function (Graur and Li 2000).
Such local duplications can cause many genes with similar function to
be clustered together.  Because there are regional biases in mutation
rate (discussed in the section on Block Structure of the Correlation
Length), these functionally-related genes will tend to have similar
mutation rates.  GO categories containing these genes will then be
biased toward the mutation rate of the region surrounding the genes.

We tested the effect of gene duplications on category rates through a
clustering analysis (see Methods). If gene duplications are not important to
category rates, genes in a hot (cold) gene category would be expected
to be distributed randomly throughout the many hot (cold) regions
around the genome, i.e. clustering of genes would be weak.  However,
if gene duplications are relevant, we would expect hot (cold) genes of
the same category to be tightly clustered since many of these genes
would have arisen by local duplications. We therefore studied the
location distribution of each of the gene categories and analyzed the
significance of its clustering, using the short-range correlation
length $\tau \sim 10^6$ base pairs (see the section on Block
Structure) as a defining length scale. This analysis was similar to
that of Williams and Hurst, who studied clustering of tissue-specific
genes (Williams and Hurst 2002), though we analyzed a larger number of
more narrowly defined gene families.

We found that some of the hot gene categories were indeed clustered,
but that none of the cold gene categories were.  The results of the
clustering for the hot and cold categories are displayed in Tables
\ref{table:GOhighsub} and \ref{table:GOlowsub}, with the clustering
p-values shown via their $-\log_{10}$ values .  Of the 21 statistically
significant hot categories, 10 categories had statistically
significant clustering ($-\log_{10} p_{cluster} > 3$).  For example,
the olfactory receptor activity category GO:0004984 has 223 genes,
with a randomly expected number of clustered genes equal to 30.6. The
actual number of clustered genes was found to be 190, which has a
p-value of less than $10^{-16}$.  In the set of 37 cold gene GO
categories, none had statistically significant clustering.  The
clustering significance is plotted versus the substitution score $z$
for all the GO categories with at least 5 members in
Fig.\ref{fig:ratecluster5}.  There were many categories of hot genes
with significant clustering ($ \log p_{cluster} > 3$), but virtually
no cold ones.

As an example of clustering in the hot gene categories, we considered
the olfactory receptors (GO:0004984). It is well-established that
olfactory receptors occur in clusters throughout the human genome
(Rouquier et al. 1998), and we likewise observed the olfactory
receptors to be highly clustered in three regions near the head,
middle, and tail of Chromosome 11 (Fig.~\ref{fig:chr11clusters}).  The
central is displayed in Fig.~\ref{fig:chr11hot}. This clustering
provided evidence that local gene duplications have influenced the
high category rate of the olfactory genes.

We next attempted to determine if the high olfactory rates are due to
a regional bias. The substitution rates of all genes are plotted in
Fig.~\ref{fig:chr11hot}, with the olfactory genes in red.  As
expected, the olfactory genes exhibited an obvious bias for higher
substitution rates than other genes.  We next calculated the
mutation rate of the region as determined from an independent measure, 
the substitution rates between ancestral repeat
sequences (green curve), using data published by Hardison et al
(Hardison et al. 2003) (see Methods).  The repeat sequence mutation
rate was notably higher in the regions where the olfactory genes
occur, showing that the hotness of the olfactory genes is a regional
property, and not specific to the genes.

Similar clustering and regional hotness were observed for other hot
gene categories. We plot the substitution rates of a cluster of
homophilic cell adhesion genes (GO:0007156) on Chromosome 5 in Fig.
\ref{fig:chr5hot}, along with the rates of nearby genes and the
ancestral repeat sequence substitution rates. The same features
observed for the olfactory genes were also present for the cell
adhesion genes: clustering, high substitution rates, and an elevated
ancestral repeat substitution rate. The repeat substitution rate
exhibited a plateau-like behavior over the region defined by the
homophilic cell adhesion genes. These factors support the
interpretation that significant numbers of hot genes have arisen by
duplications in inherently hot regions of the genome.

\subsection{Block Structure of the Substitution Rate}

Several explanations have been proposed that could account for the
regional biases in mutation rate (Mouse Genome Sequencing Consortium
2002), including recombination-associated mutagenesis (Lercher and
Hurst 2002, Perry and Ashworth 1999), strand asymmetry in mutation
rates (Francino 1997) and inhomogeneous timing of DNA replication
(Wolfe et al. 1989, Gu and Li 1994).  The structure of regional biases
could be considered from the perspective of amino-acid changing
substitutions as well, since linked proteins have been known to have
similar substitution rates (Williams and Hurst 2000, Williams and
Hurst 2002). However, the silent sites may be easier to comprehend,
since protein sequences are more likely to be complicated by
non-neutral pressures.

To shed light on the structural properties of the hot and cold
mutational regions, we measured the length scale over which
substitution rates are correlated.  Previously, correlations have been
observed in blocks of particular physical ($5$ megabases) (Hardison et
al. 2003) or genetic ($1,2,5,200$ centimorgans) (Matassi et al.  1999,
Lercher et al. 2001) size.  While these studies have focused on
whether correlations exist at certain length scales, it is informative
to measure the decay of correlations with distance.  We therefore
measured the length scale of substitution rate correlation, using an
analysis of the correlation function (Huang 1987)
\begin{equation}
<r(0)r(t)>,
\end{equation}
where $r(t)$ is the substitution rate of a gene $t$ basepairs
downstream of a gene with substitution rate $r(0)$, and $<...>$
indicates an average over the available data (see Methods).  We expect
that at small $t$, the correlation function will be positive and then
decrease with $t$ as rates become decoupled.  The length scale over
which this decay occurs serves as a measure of the typical size of hot
or cold regions. The rate correlation function is plotted in
Fig.~\ref{fig:rate_correlation} versus both the human and mouse values
for $t$.

We observed two notable behaviors: 1) a strong correlation which
decays over a region of approximately 1 megabase, and 2) a longer
range correlation which plateaus over a region of approximately 10
megabases.  At larger distances, correlations are weaker. For example,
the human curve first dips below the $<r(0)r(t)> = 0$ threshold at
approximately 11 Mb and the mouse curve first crosses it at
approximately 9 Mb. This suggests that there are multiple phenomena
which control the mutation rate of regions, both long (10 Mb) and
short (1 Mb) length scale.

We also measured the characteristic short-range correlation length
using an exponential fit. The correlation length $\tau$ was determined
by fitting the data to the functional form
\begin{equation}
<r(0)r(t)> = A_0 \exp(-t/\tau) + A_\infty,
\end{equation}
where $A_\infty$ is the correlation at long distances and
$(A_0+A_\infty)$ is the correlation at zero distance.  Because of the
observed plateauing behavior of the data, we performed our curve fit
over the region $t \in [0,10000000]$. For the human data we obtained
$A_0 = 0.83, \tau = 1.21 \times 10^6, A_\infty = 0.39$.  For the mouse
data we found values of a similar magnitude ($A_0 = 1.08, \tau = 0.73
\times 10^6, A_\infty = 0.32$) suggesting that short range mutational
processes may be alike in mouse and human.  The long range correlation
$A_\infty$ was at least an order of magnitude larger than would be
expected by chance at all distances up to 10 megabases (see Methods).

It is unclear what factors are responsible for these two length scales
of rate correlation, though some guesses are possible.  For the short
range effect, one process that occurs on the appropriate lengthscale
is DNA replication (Alberts et al. 1994).  Replication origins in a
concerted unit activate under similar timing, similar cell conditions,
and could have a common regulatory mechanism, making it a reasonable
to expect the DNA in such a unit to have similar mutation rates.

Long range correlations have previously been observed at
chromosomal-size distances in particular regions of the genome, e.g.
it is known that Chromosome 19 is generally hotter than other
chromosomes (Castresana 2002, Lercher et al. 2001). However, the 10
megabase correlation was not simply due to selection on chromosomes.
We removed the respective chromosomal average from each substitution
rate and repeated the correlation analysis, finding that $A_\infty$
retained a significant value of $\sim 0.2$.  One possible mechanistic
explanation for the long-range correlation is suggested by the finding
of Lercher et al that recombination rate and substitution rate are
correlated even in blocks extending to 30 megabases (Lercher and Hurst
2002).  Therefore, if large regions of similar recombination rate
exist, they could be related to the long-range 4-fold correlation
effects we observed.

\section{Discussion}

\subsection{Evidence for Selection}

Recently, there has been evidence for selective factors influencing
gene location in yeast (Pal and Hurst 2003). This suggests the
possibility that similar phenomena affect gene locations in
mouse/human as well.  We therefore considered whether regional
mutation rates could have selectively influenced the types of genes
occurring in different loci in the genome. Selection due to the local
mutation rate would require different mechanisms than that observable
through the traditional measure $K_A/K_S$, which quantifies selection
on point mutations.  For example, regional mutation rates could have
influenced the fitness of the genome after events that cause gene 
relocation, such as gene transposition or chromosomal recombination. 
Or perhaps the duplication of certain genes provided a fitness benefit 
(a mechanism
possibly relevant for the hot clustered categories). Differential
duplication rates could force a category to have a mutation rate bias,
due to the block structure of the mutation rate and the fact that
duplications occur locally.

The observed categories of hot and cold genes suggest gene locations
have been selectively influenced by regional mutation rates.  This is
because if mutation rates were unrelated to gene function, then the
lists of hot and cold categories would be expected to be random, i.e.
the lists shown in Tables \ref{table:GOhighsub} and
\ref{table:GOlowsub} would have been evenly sampled from all possible
Gene Ontology categories.  However, this was not the case, as the hot
and cold categories each had strong internal commonalities.

The hot categories were found to be biased toward receptor activities
or roles in extracellular communication. Intriguingly, arguments based
on protein-level effects appear applicable to the silent-site hotness
of these categories. Cellular receptors and those involved in
extracellular communication are the proteins that most directly
interact with the environment, and are therefore the most likely to
have experienced a dynamically changing set of selection pressures.
This variability of selection pressures would have made it favorable
for them to be in hot regions, in order that new mutations be possible
to deal with new stimuli. Examples of hot categories with known
protein-level diversification pressures include the olfactory
receptors (Lane et al. 2001), immune genes (Papavasiliou and Schatz
2002), and cell adhesion genes (Uemura 1998, Tasic et al. 2002).

Arguments normally applied to protein-level selection were found to be
appropriate for cold mutation rate categories as well. Cold categories
were often related to transcription or other regulatory processes.
Regulatory proteins should be tuned to interact with many different
nucleic acid or protein targets, in contrast with receptor proteins
which typically interact with only a particular ligand.  Mutations to
regulatory proteins would therefore be expected to be more
deleterious, and hence it would be beneficial for them to have low
mutation rates. Strong conservation pressures in the cold categories
could also be related to their roles as housekeeping genes (Zhang and
Li 2003), or as essential genes. For example, in the dataset of
(Winzeler et al. 1999), 81 out of 356 essential yeast genes were
involved in transcription, whereas only 4 were involved in signal
transduction, the function most similar to extracellular communication
for which data was available.

The applicability of protein-level arguments to synonymous mutation
rates suggests that $K_S$ and $K_A$ are under similar pressures. This
is consistent with what would be expected if gene locations have
evolved to make use of the block structure of the mutation rate, since
relocation to a hot (cold) spot would increase propensities for both
high (low) $K_A$ and $K_S$. More quantitatively, we observed that
$K_S$ category biases were similar to category biases caused by
selection on amino acid changing point substitutions -- i.e. selection
observable through the ratio $K_A/K_S$. We performed a Gene Ontology
$z$-score analysis on $K_A/K_S$ (for consistency, the CODEML method in
PAML was used to calculate both $K_A$ and $K_S$).  There were 8 hot
categories common to both the 4-fold and $K_A/K_S$ classifications (
immune response, proteolysis and peptidolysis receptor activity,
integral to membrane, chymotrypsin activity, cell adhesion, trypsin
activity, olfactory receptor activity), and 17 common cold categories
(nucleus, regulation of transcription, transcription factor activity,
RNA binding activity, development, ribonucleoprotein complex,
ribonucleoprotein complex, protein transporter activity, protein
serine/threonine kinase activity, ubiquitin conjugating enzyme
activity, GTP binding activity, ubiquitin-dependent protein
catabolism, translation regulator activity, intracellular protein
transport, neurogenesis, ubiquitin cycle, cytoplasm, regulation of
transcription from Pol II promoter).  The strong commonalities between
the two types of classification suggest that the selective forces that
influenced amino-acid changing point mutations also influenced gene
locations. The hot and cold categories derived from $K_A/K_S$ are
available in the supplementary materials.

Selection on gene locations would provide an evolutionary explanation
for the puzzle of why $K_A$ and $K_S$ are correlated beyond levels
expected by neutral evolutionary theory (Ohta and Ina 1995, Mouchiroud
et al. 1995).  Assuming 4-fold sites are neutral, locational selection
would have to be realized through the influence of the local mutation
rate $K_S$ on the amino-acid changing mutation rate $K_A$. Thus
locational selection and point mutation-based amino acid selection
would behave similarly with respect to positive or negative selection
on protein sequence, increasing the correlation of $K_A$ and $K_S$,
even if mutations to any individual 4-fold site did not provide a
fitness benefit.

One caveat is that other, not necessarily exclusive, explanations for
the strong correlation of $K_A$ and $K_S$ have been proposed as well
-- most notably simultaneous substitutions at adjacent sites, so
called tandem substitutions (Smith and Hurst 1999b).  Tandem
substitutions were not sufficient to explain our hot and cold
categories, however. We rederived sets of hot and cold categories
after correcting for tandem effects (see Methods) and once again found
similar results. For example, the 6 hottest categories (of 22
significant) were dynein ATPase activity, receptor activity,
homophilic cell adhesion, olfactory receptor activity, integral to
membrane, and calcium ion binding activity. The 6 coldest (of 36) were
nucleus, regulation of transcription DNA-dependent, RNA binding
activity, transcription factor activity, development, and
ribonucleoprotein complex.

\subsection{Mechanisms}

For the hot clustered categories, it may be that high mutation rates
and high rates of gene duplication are tied to a hidden variable which
imposes both phenomena simultaneously. One possibility is the
recombination rate along the genome, which Pal and Hurst found to have
selective effects in yeast (Pal and Hurst 2003). For example, two
mechanisms for diversification, gene duplication and mutation, can
both be accelerated by recombination (Lercher and Hurst 2002, Graur
and Li 2000).  High recombination rates are relevant for a number of
the hot gene categories we have studied, as they have been suggested
for the protocadherins (Wu et al.  2001), immune response
(Papavasiliou and Schatz 2002), and olfactory families (Sharon et al.
1999).  Because both gene duplication and point mutation are useful
for diversifying a family, it is difficult to separate the
significance of mutation rate and recombination rate. Pal and Hurst
offered preliminary evidence that in yeast, selection acts on the
recombination rate, but not point mutation rates (Pal and Hurst 2003).
However, we have observed unusual rate biases for non-clustered gene
categories as well, for which recombination would not be expected to play 
a role.

Cold gene categories are not clustered; therefore, the existence of
cold categories (as well as non-clustered hot categories) can not be
attributed to duplication events. One alternate phenomenon that could
cause cold category biases is gene relocation to cold regions.  The
concept of relocation brings up a number of questions.  First, if cold
genes have relocated, this leaves one wondering in what sort of
environment cold genes originated. One speculative possibility is that
these genes developed in regions of high recombination (the hot
regions) which would have allowed for fast duplication and functional
diversification, and later dispersed to cooler regions as their
functions became fixed.  Second, it is unclear whether gene
relocations occur frequently enough to account for the observed rate
biases.  This issue is complicated by the fact that genes have arisen
at different times. Many of the cold gene categories occur in diverse
sets of tissues and have important regulatory effects, suggesting they
should be relatively old. This old age may have allowed them enough
time to redistribute through the genome.

We verified the correlations of substitution rates along the genome
and showed that these correlations lead to an excess of hot and cold
genes, confirming studies by Matassi et al (Matassi et al. 1999) and
Hardison et al (Hardison et al. 2003).  Our results appear to disagree
with those of Kumar and Subramanian (Kumar and Subramanian 2002), who
reported that mutation rates are uniform in the genome. While our rate
measurements were qualitatively similar to those of Kumar and
Subramanian, one beneficial addition we made was the use of a
normalized rate which accounts for the length-dependence of rate
variance, allowing genes of differing lengths to be treated equally in
Fig.~\ref{fig:mutdistribution}. Our correlation length analysis
revealed two scales of rate correlation: a short decay length of 1
megabase and a long range length extending along a syntenous block up
to distances of 10 megabases.  We have very speculatively proposed
that DNA replication units and DNA recombination may be relevant to
these length scales. More generally, it is hoped that these scale
determinations will be helpful in placing constraints on possible
processes that control mutation rate.

Some data issues suggest topics for further exploration.  First, the
resolution of our analysis is dictated by the structure of the Gene
Ontology, which currently has 16000 categories, but is evolving. Our
category inferences should become more specific as GO gene assignments
improve.  Second, multispecies data will be invaluable in revealing
the mutations that have occurred in each lineage. One promising early
result from human-chimpanzee comparisons, based on a set of 96
orthologs derived from HOVERGEN release 44 (Duret et al. 1994), is
that olfactory receptors are a hot category. Unfortunately, this is
the only statistically significant hot or cold category at present,
due to the lack of data. However, inferences should improve rapidly as
more chimpanzee gene identifications become available.

\section{Methods}

\subsection{Ortholog Generation}

We downloaded a list of the available 37347 human and 27504 mouse
peptides from the Ensembl sequence database (www.ensembl.org), then
used BLAST (Altschul et al. 1990) to find orthologous peptide
sequences between the genomes. The peptides studied were the set of
all known or predicted peptides in the Ensembl 12.31.1 human and
12.3.1 mouse datasets.  Sequences were designated as orthologous if
the two peptides were each other's mutual best hit in the opposing
databases, as determined by blastall, and the E-value for the match
was $10^{-10}$ (using the higher score as a worst case bound) or
better. We chose this method of ortholog determination to get a
one-to-one relationship between proteins.  We found 14790 ortholog
pairs, a coverage rate of approximately 50\% in mouse and 40 \% in
human.  The observed E-values between orthologs have a median value of
0.0 ($<$ 1e-180).  The aligned peptide orthologs were then used in
conjunction with Ensembl CDNA data to determine aligned orthologous
CDNA. For the chimp-human comparison, human genes from ENSEMBL were
compared to chimp genes from HOVERGEN. A mutual best hit criterion was
used to determine the set of 96 orthologs.

We manually inspected the mouse-human synteny of the olfactory gene
cluster of Fig.~\ref{fig:chr11hot} to verify that orthologs were
assigned correctly. This was to address the concern that orthologs are
more difficult to assign in gene categories with many homologous 
members, since incorrect assignments could distort substitution rates.  
The synteny structure  was found to be almost totally conserved for 
these genes, as it was for the cell adhesion genes in Fig.~\ref{fig:chr5hot}.

\subsection{Calculation of Substitution Rates}

We calculated the distribution of substitution rates between the mouse
and human genomes using the 4fold sites of orthologous genes.  4fold
sites are the 3rd bases of codons for which the amino acid is
specified by the first two bases. For each of the orthologous gene
pairs, we calculated $p$, the fraction of 4fold sites in which the
mouse base differs from the human base.  The average value of $p$ over
all 4-fold sites in all orthologs was $<p> = 0.337.$ The average 4-fold
substitution rate on a genewise basis was 0.338 with a standard
deviation of 0.080.  These rates were in agreement with substitution
rates measured in other studies of 4-fold sites or in ancestral
repeats (Hardison et al. 2003, Mouse Genome Sequencing Consortium
2002).

Because genes are of finite length, stochastic effects can cause
substitution rates to vary from gene to gene, even if all 4-fold sites
mutate at the same rate.  We defined a normalized substitution rate
to correct for these finite-size effects.  A gene with $N$ 4fold
sites was modeled as having $N$ independent events in which
substitution can occur with probability $<p>$. This formulation can
fit both the Jukes-Cantor one-parameter or the Kimura two-parameter
model for mutation matrices (Durbin et al. 1998). Although this model
is not as sophisticated as other more modern treatments (for example,
see (Tavere 1986, Tamura and Nei 1993, Li 1993, Goldman and Yang
1994)), it gives an easily falsifiable prediction that the rate
distribution, in the absence of regional correlations, is Normal, due
to the Central Limit Theorem (Rice 1995).

\begin{figure}\caption{\ }\label{fig:mutdistribution}
\begin{center}
  \includegraphics[width=3.0in]{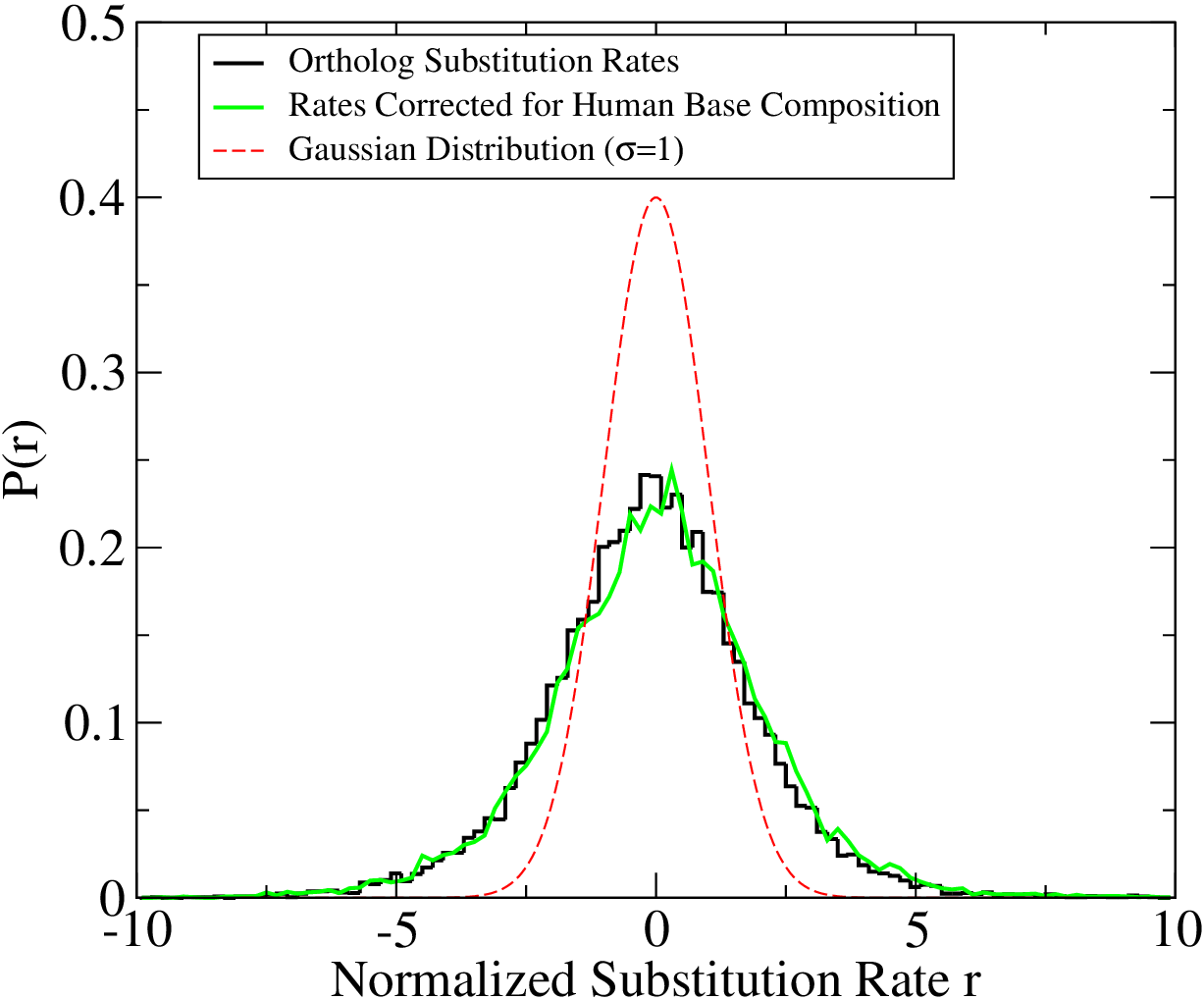}
\end{center}
\end{figure}

\begin{figure}\caption{\ }\label{fig:ratecluster5}
\begin{center}
  \includegraphics[width=3.0in]{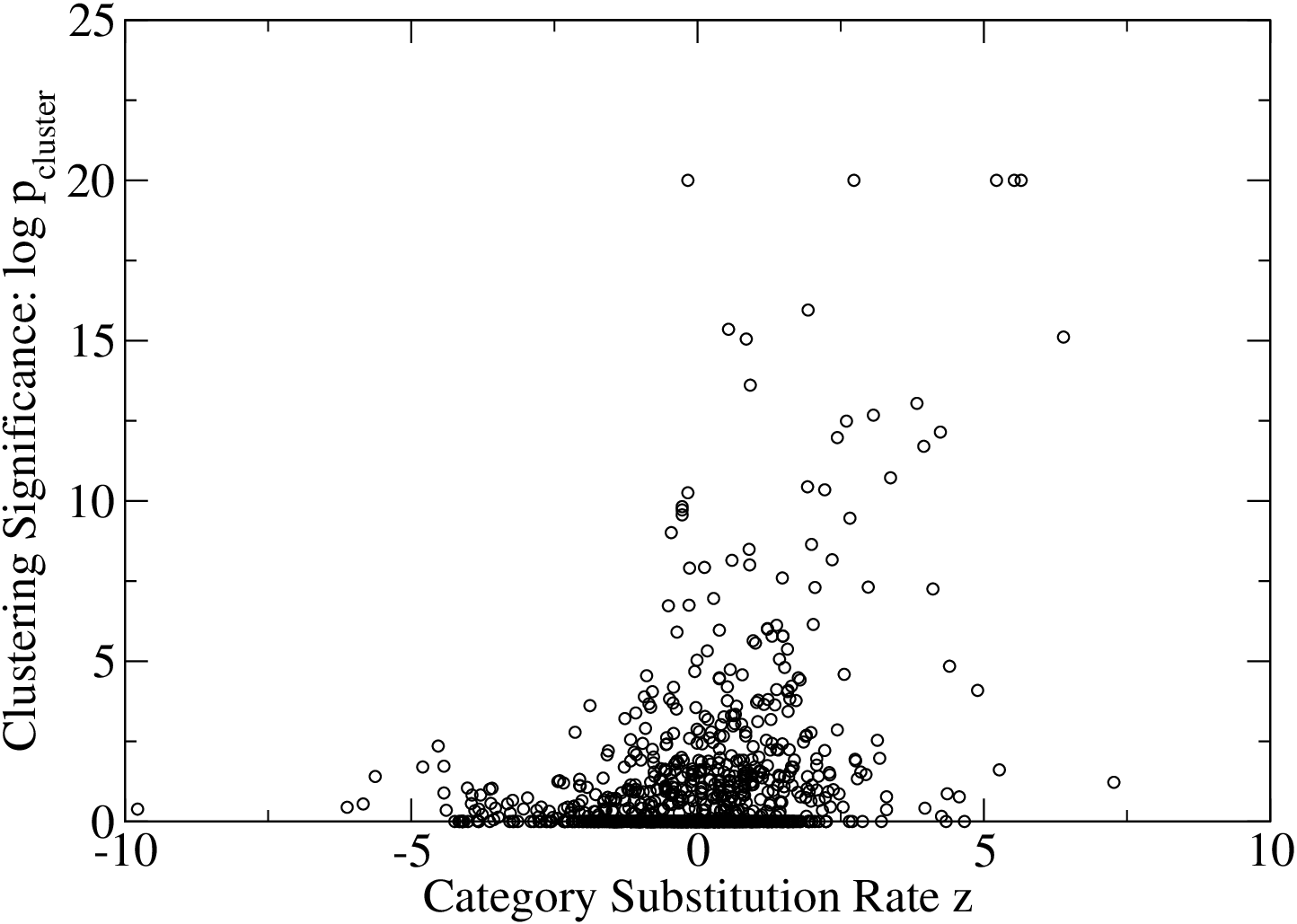}
\end{center}
\end{figure}

\begin{figure}\caption{\ }\label{fig:chr11clusters}
\begin{center}
  \includegraphics[width=3.0in]{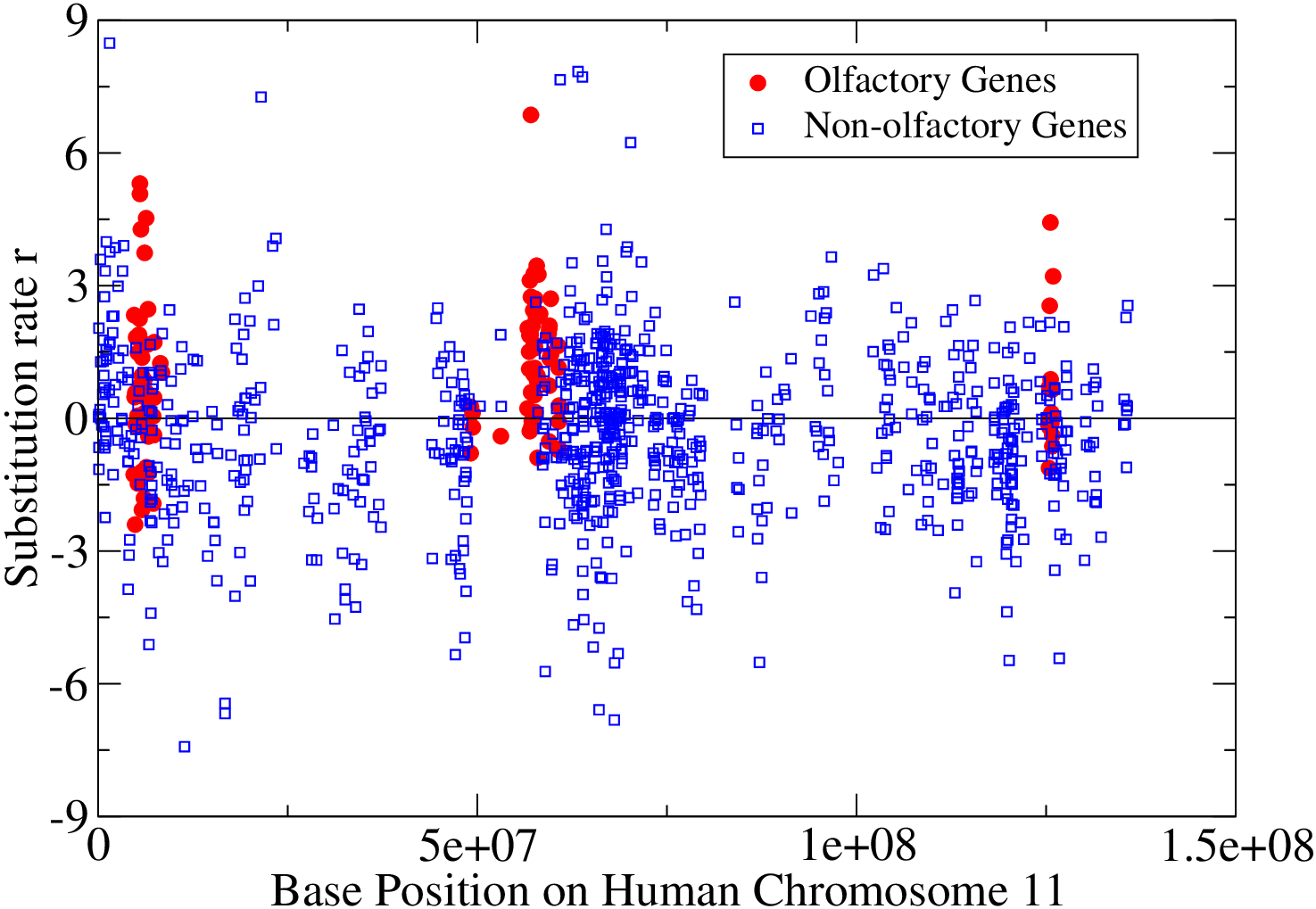}
\end{center}
\end{figure}

\begin{figure}\caption{\ }\label{fig:chr11hot}
\begin{center}
  \includegraphics[width=3.0in]{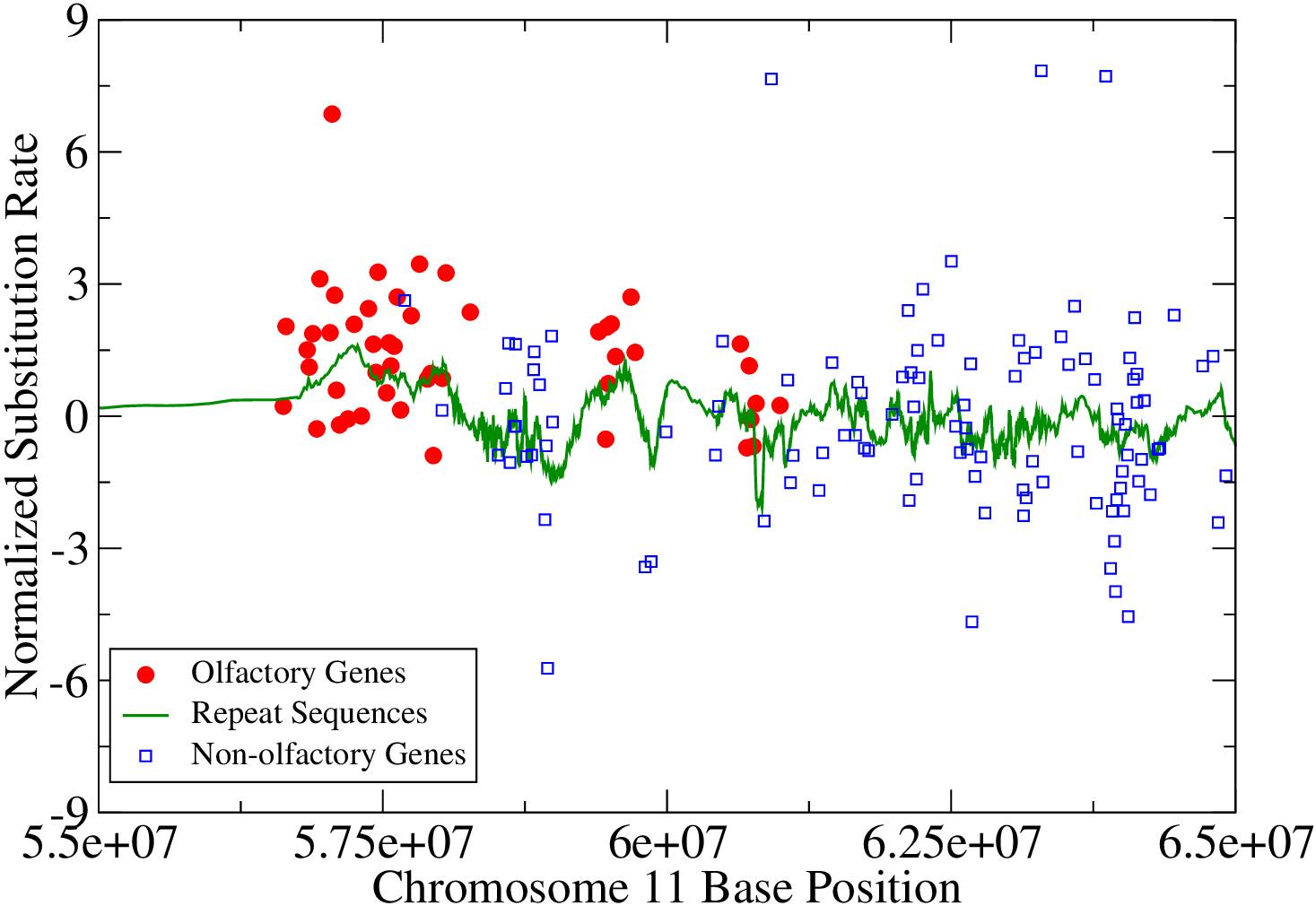}
\end{center}
\end{figure}

\begin{figure}\caption{\ }\label{fig:chr5hot}
\begin{center}
  \includegraphics[width=3.0in]{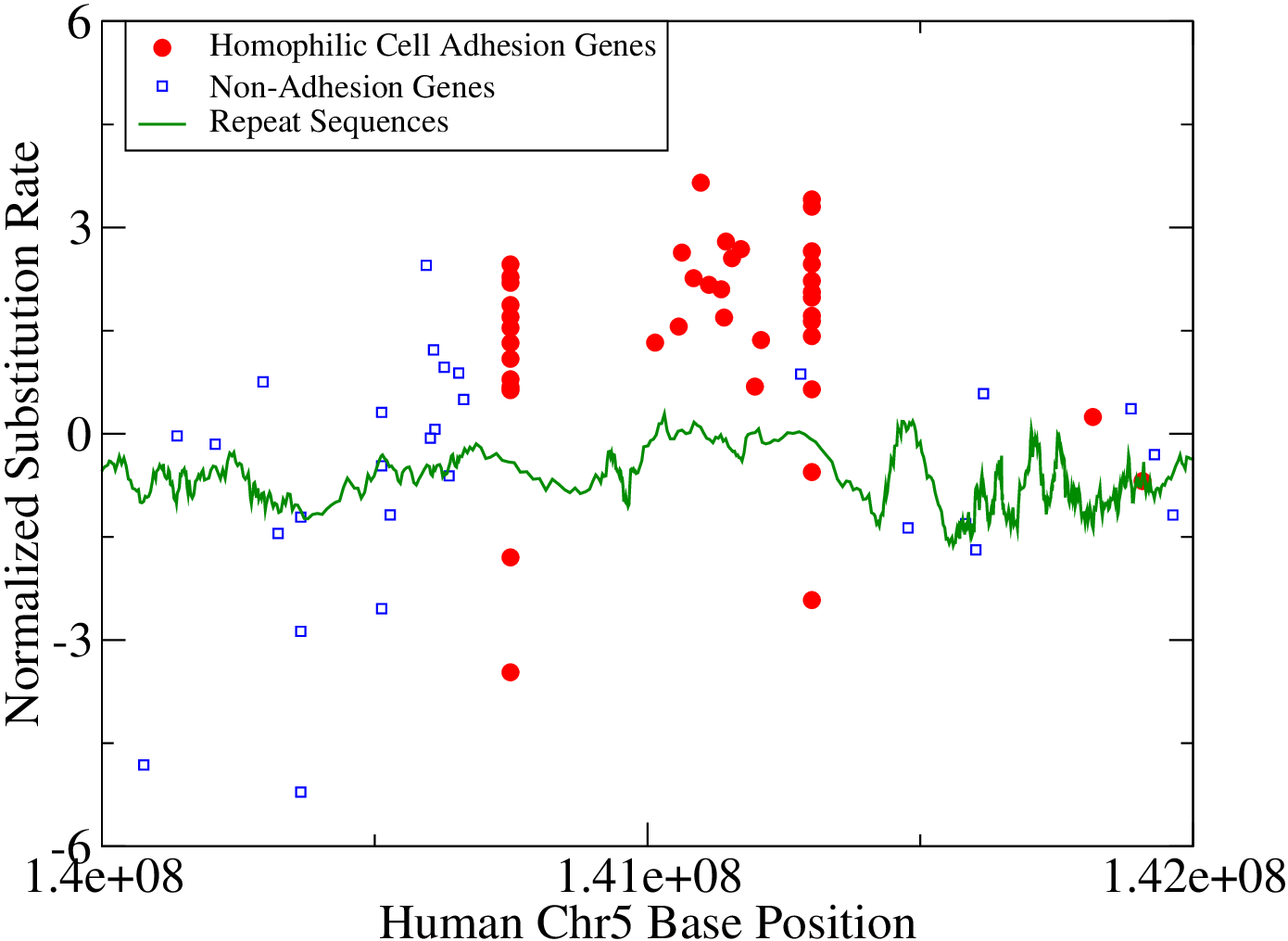}
\end{center}
\end{figure}

Under this model, at each $N$ the distribution of substitution rates 
can be described by a binomial distribution with standard deviation 
$\sigma(N) = \sqrt{<p>(1-<p>)/N}$.  Therefore gene substitution rates were 
normalized by their respective $\sigma(N)$ to get one universal rate 
distribution, which in the limit of many data points should approach the Normal
distribution ($2\pi)^{-1/2} \exp(-x^2/2$).  We defined the normalized
substitution rate to be
\begin{equation}
r \equiv (p-<p>)/\sigma(N), 
\end{equation}
where $p$ is the actual 4fold substitution rate in the gene. The
values of $r$ for all ortholog pairs were used to calculate the
distribution shown in Fig.~\ref{fig:mutdistribution}.

The actual rate distribution in genes was found to be skewed toward
high or low mutation rates, as shown in
Fig.~\ref{fig:mutdistribution}.  The observed distribution had a
standard deviation of 2.04, significantly higher than the expected
$\sigma = 1$. Similar excesses of hot and cold genes
were found even when corrections were made for base composition. To
verify this, we calculated a normalized mutation rate using a
4-parameter model in which each site of type A, C, G, or T in the
human sequence has its own substitution probability.  For each human
base (A,C,G,T) we measured the substitution rate at the corresponding
4-fold locations, yielding 4 rates $<p_A>,<p_C>,<p_G>,<p_T>$. Based on
these rates, we then calculated the expected frequency and variance of
substitutions for a gene given the gene's base composition at the
4-fold sites. This yielded a distribution nearly identical to that in
the 1-parameter model (Fig.~\ref{fig:mutdistribution}).
   
We also tested whether neighboring genes have similar substitution
rates.  The orthologs were ordered by their location along the human
genome, after which we calculated the Pearson correlation of a gene's
substitution rate $r$ with that of its following gene. We used only
neighboring genes which were in syntenous blocks as defined by all
three conditions of monotonicity (the genes are ordered the same in
both species), consistent strand orientation (a block is either in the
same strand orientation in both species or completely reversed), and
consistent chromosome (no chromosome changes in either
species in a block), yielding a dataset of 11087 neighbor pairs.
Under this condition, the Pearson correlation was 0.26, corresponding
to a highly significant p-value of $10^{-189}$.

\subsection{$z$-score calculation for GO categories}

For each GO category, we calculated a normalized substitution rate
($z$-score) based on the substitution rates of all members of that
category. 9966 of the genes in our ortholog set had Gene Ontology
classifications available. The $z$-score was defined to be
\begin{equation}
z \equiv \frac{<r>_{GO} - <r>_{all}}{\sigma_{all}/\sqrt{N_{GO}}},
\end{equation}
where $<r>_{GO}$ is the average substitution rate $r$ for the genes in
the GO category, $<r>_{all}$ is the average $r$ for all of the genes
with Gene Ontology classifications, $\sigma_{all}$ is the genewise
standard deviation, and $N_{GO}$ is the number of genes in the
category. The p-value for $z$ was determined from the probability that
a gaussian-distributed variable (again using the Central Limit
Theorem) takes on a value $\ge z$.  To reduce the problem of outliers,
we limited our analysis to the GO categories containing at least 5
genes, of which there are 997, and accordingly set a p-value cutoff of
$1/997 \sim 10^{-3}$. We expressed the significance in terms of
$-\log_{10} p_{z},$ which should have a value larger than $3$ to be
statistically significant.

$z$-scores corrected for tandem substitutions were calculated by first
removing all possible tandem substitution sites from the dataset. That
is, 4-fold sites were only accepted into the dataset if both the
preceding and following bases matched in the two species. After
culling the dataset, we calculated rates and category $z$-scores as
before.

\subsection{Clustering Analysis}
To measure clustering, for each gene in a GO category we tested
whether it had another category member downstream of it within the
short-range correlation length $\tau = 10^6$ basepairs.  In each GO
category, we calculated the number of genes satisfying this condition,
defining this to be the number of ``clustered genes.''  This
``downstream'' criterion (rather than a symmetric one) was used to
avoid the problem of double counting of genes when several are close
together. To test the statistical significance of the number of
clustered genes in a category, we used bootstrapping. For each GO
category, we performed 5000 random trials of selecting $N_{GO}$ random
genes from the entire set of orthologs, where $N_{GO}$ is the number
of genes in the GO category. In each trial, we counted the number of
clustered genes in this randomly selected group.  The average number
of clustered genes was used to approximate the random number of
clustered genes by a Poisson distribution.  These Poisson statistics
were then used to calculate the significance of the number of
clustered genes for the GO category.  A poisson distribution is
appropriate as long as clustering of neighbors is a rare event, i.e.
as long as $N_{GO} \ll N_{all genes}$, which was generally the case.
The random distributions were visually inspected and found to agree
with the shape of the Poisson curve. To generate the data for Tables 1
and 2 we also limited ourselves to the 997 categories with at least 5
genes, implying that $-\log_{10} p_{cluster} > 3$ is the cutoff for
significance.

\subsection{Calculation of Repeat Sequence Mutation Rates}

Aligned repeat sequences between mouse and human were obtained from
the dataset of Hardison et al (Hardison et al.  2003).  For each
repeat, positions in which a base was defined for both the mouse and
human sequence were used to calculate a normalized substitution rate,
in analogy with the method used for the 4-fold sites.  The genome-wide
average value of $p$ in these repeat sequences was 0.33, which was
very close to the value for 4-fold sites 0.34. The start position of
each repeat sequence was used to define its location in the genome. In
order to determine the locations of repeat sequences (based on the
June 2002 UCSC genome map) along the physical map used for the gene
sequences (based on the Ensembl May 2003 map), gene locations
according to the two maps were compared. Repeat sequence
locations were then corrected using the location differences of
nearby genes.  For clarity, the ancestral repeat values shown in
Fig.~\ref{fig:chr11hot} and Fig.~\ref{fig:chr5hot} were smoothed using
a moving-window average of 20 repeat sequences.

\subsection{Correlation Length Calculation}
\begin{figure}\caption{\ }\label{fig:rate_correlation}
\begin{center}
  \includegraphics[width=3.0in]{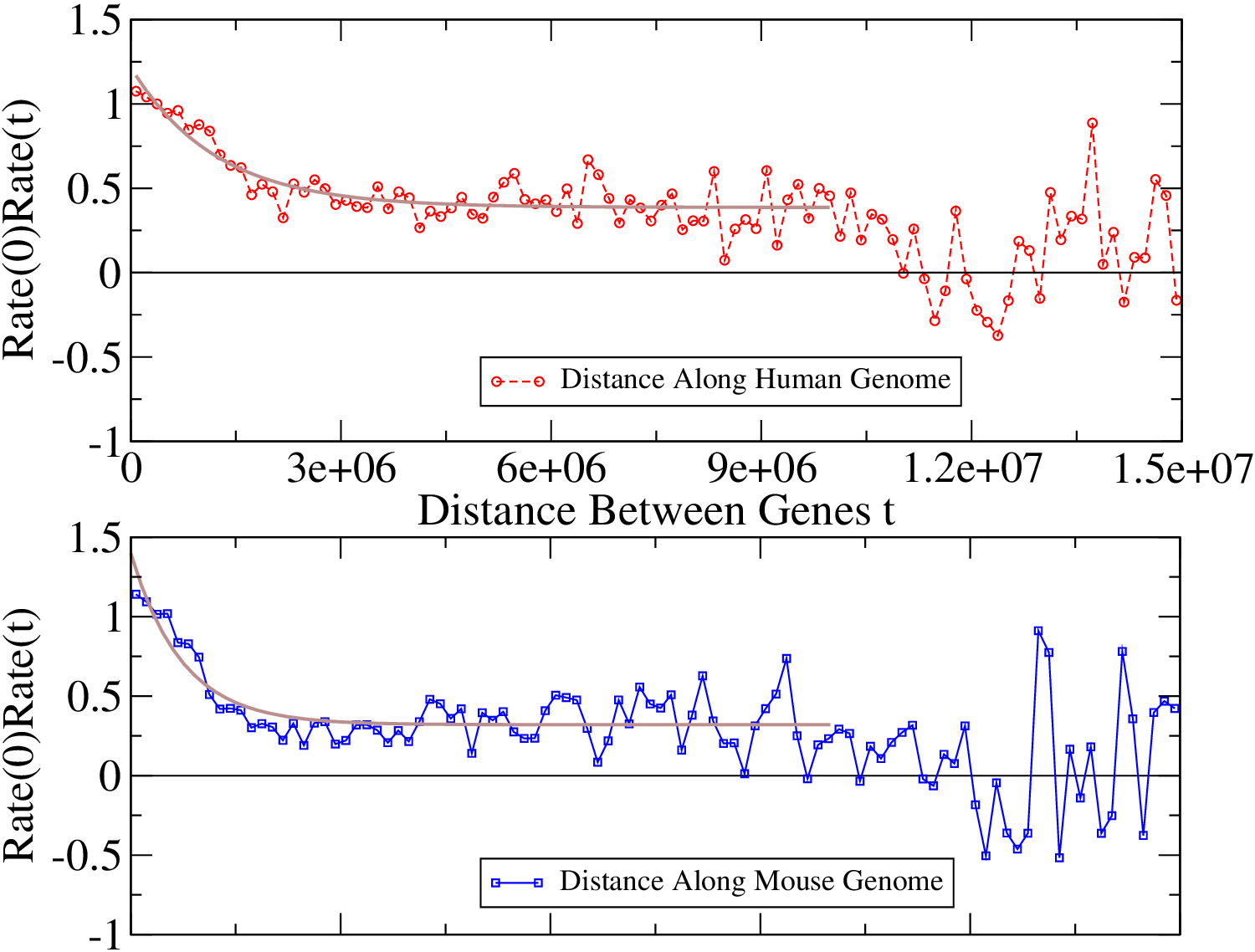}
\end{center}
\end{figure}

We considered all pairs of genes on continuous orthologous blocks,
starting from the first neighbor up to the 35th gene downstream.  This
allowed us to get hundreds of measurements of $r(0)r(t)$ for $t$
values even as large as several megabases. We binned these data into
100 uniformly spaced groups covering $t\in[0,15000000]$ and then
averaged over each of these bins to determine the correlation function
$<r(0)r(t)>$. The data were plentiful enough for the averaged values
shown in Fig.~\ref{fig:rate_correlation} to be statistically
significant. It was difficult to extend to larger values of $t$ since
the amount of data decreases with $t$, a fact manifested in the
increasing fluctuations at larger $t$ in
Fig.~\ref{fig:rate_correlation}.  For example, the value of the
average correlation $<r(0)r(t)>$ at $t=15$ megabases in the human data
of Fig.~\ref{fig:rate_correlation} was based on only 79 measurements,
whereas at $t=75000$ it was based on 22860 measurements.  For genes
with alternative splicings, only one of the genes was used, in order
to avoid spurious effects caused by reuse of DNA.  Orthologous block
boundaries were defined by genes at which the chromosome changes in
either species. Monotonicity and consistent strand orientation were
ignored in order to obtain blocks with large values of $t$.  Most of
the $r(0)r(t)$ data comes from blocks at least several megabases long.
Approximately 5\% is in blocks of size less than $10^6$ base pairs,
55\% is in blocks of size between $10^6$ and $10^7$ base pairs, and
the remaining 40\% is in larger blocks.

The long-range correlation shown in Fig.~\ref{fig:rate_correlation}
was statistically significant. Theoretically, fluctuations in
$<r(i)r(j)>$ should be of order $\sim O(1/\sqrt{N})$, where $N$ is the
number of data samples in a bin. At a distance of 10 megabases, there
were $\sim 400$ samples, corresponding to an uncertainty of $\sim
0.05$. This uncertainty was an order of magnitude smaller than the
observed value of $A_\infty = 0.4$.

\section{Acknowledgments}
This material is based upon work supported by the National Science
Foundation under a grant awarded in 2003. Any opinions, findings, and
conclusions or recommendations expressed in this publication are those
of the authors and do not necessarily reflect the views of the
National Science Foundation. JC would like to thank T. Hwa, D. Petrov,
and C. S. Chin for comments on the manuscript.

\pagebreak

\section{Figure Captions}

\medskip
\noindent
{\bf Figure 1. Distribution of normalized substitution rates}
{Histogram of substitution rates based on 14790 orthologous mouse and
  human genes (black curve). The rate distribution has significantly
  more genes at high and low rates than the expected Normal
  distribution (red curve). This bias toward high and low rates
  remains even when rates are corrected for human base composition
  (green curve).}  \medskip
\noindent
{\bf Figure 2. Clustering versus substitution rate for GO categories
  containing at least 5 members} {Virtually all clustered gene
  categories have higher than average substitution rates ($z > 0$). }

\medskip
\noindent
{\bf Figure 3. Clustering of olfactory genes on human Chromosome 11} {
  The olfactory genes are clustered into three regions along the
  chromosome.  The substitution rates of the olfactory genes are
  almost all hot, while the non-olfactory genes are distributed around
  $r=0$.}

\medskip
\noindent
{\bf Figure 4. Olfactory genes lie in a mutational hot spot} {
  Substitution rates of the olfactory genes in the central region of
  human Chromosome 11. The substitution rate of ancestral repeat
  sequences is higher in the region where the olfactory genes lie.  }

\medskip
\noindent
{\bf Figure 5. Homophilic cell adhesion genes also lie in a hot spot}
{ Substitution rates of a cluster of homophilic cell adhesion genes on
  human Chromosome 5, along with substitution rates of other genes and
  ancestral repeat sequences. The repeat sequence substitution rate
  plateaus at a higher level in this region. }

\medskip
\noindent
{\bf Figure 6. Correlation length analysis of substitution rates} {
  Correlation of substitution rates in syntenous blocks as a function
  of distance between genes measured along the human chromosome (top)
  and measured along the mouse chromosome (bottom).  There are two
  length scales of correlation decay: a short one of one megabase and
  a long one of 10 megabases.  The curve fits are for $<r(0)r(t)> =
  A_0 \exp(-t/\tau) + A_\infty$ for the region $t \in [0,10000000]$.}

\medskip
\noindent
{\bf Table 1. Statistically significant hot Gene Ontology categories}
{ Listed are the categories with $z \ge 0$ having at least 5 genes and
  $p_z \le 10^{-3}$, sorted by statistical significance ($-\log_{10}
  p_z$).  There is a bias toward proteins involved in extra-cellular
  communication.  Several of the categories have an unusual number of
  clustered genes ($-\log_{10} p_{cluster} \ge 3$).}

\medskip
\noindent
{\bf Table 2.  Statistically significant cold Gene Ontology
  categories} {Listed are the categories with $z \le 0$ having at
  least 5 genes and $p_z \le 10^{-3}$, sorted by statistical
  significance. There is a bias toward proteins involved in DNA, RNA,
  or protein regulation.  None of the cold categories have
  statistically significant clustering.}

\pagebreak

\bigskip

\bigskip

\bigskip

\bigskip

\bigskip

\bigskip

\begin{table}\caption{\ }\label{table:GOhighsub}\end{table}
\begin{tabular}{ccccccc}
\hline
GO Id&Genes&Clustered Genes&$-\log_{10} p_{cluster}$&$z$&$-\log_{10} p_{z}$&Description\\
\hline
GO:0008567&10&1&1.22&  7.27& 12.44&dynein ATPase activity\\
GO:0004872&634&327&15.11&  6.39&  9.77&receptor activity\\
GO:0004984&223&190&83.14&  5.65&  7.79&olfactory receptor activity\\
GO:0007186&502&276&25.72&  5.53&  7.49&G-protein coupled receptor protein signaling pathway\\
GO:0016021&1401&794&1.61&  5.27&  6.86&integral to membrane\\
GO:0007156&90&47&27.97&  5.22&  6.76&homophilic cell adhesion\\
GO:0005509&449&153&4.09&  4.89&  6.01&calcium ion binding activity\\
GO:0030286&12&0&0.00&  4.66&  5.50&dynein complex\\
GO:0008152&250&44&0.77&  4.57&  5.31&metabolism\\
GO:0006508&355&111&4.84&  4.40&  4.96&proteolysis and peptidolysis\\
GO:0007018&31&2&0.87&  4.36&  4.88&microtubule-based movement\\
GO:0015029&6&0&0.00&  4.34&  4.85&internalization receptor activity\\
GO:0003824&133&10&0.16&  4.26&  4.68&enzyme activity\\
GO:0004263&62&21&12.15&  4.24&  4.66&chymotrypsin activity\\
GO:0007155&272&84&7.25&  4.11&  4.39&cell adhesion\\
GO:0003777&27&1&0.41&  3.97&  4.15&microtubule motor activity\\
GO:0004295&72&23&11.71&  3.95&  4.11&trypsin activity\\
GO:0008014&9&7&13.04&  3.83&  3.90&calcium-dependent cell adhesion molecule activity\\
GO:0006955&282&100&10.72&  3.37&  3.12&immune response\\
GO:0016020&577&175&0.37&  3.30&  3.02&membrane\\
GO:0005975&148&18&0.77&  3.30&  3.01&carbohydrate metabolism\\
\end{tabular}
\bigskip

\begin{table}\caption{\ }\label{table:GOlowsub}\end{table}
\begin{tabular}{ccccccc}
\hline
GO Id&Genes&Clustered Genes&$-\log_{10} p_{cluster}$&$z$&$-\log_{10} p_{z}$&Description\\
\hline
GO:0005634&1741&1032&0.39& -9.78& 21.85&nucleus\\
GO:0006355&910&380&0.44& -6.12&  9.04&regulation of transcription, DNA-dependent\\
GO:0003700&546&164&0.55& -5.84&  8.30&transcription factor activity\\
GO:0003723&298&66&1.41& -5.63&  7.75&RNA binding activity\\
GO:0007275&367&96&1.70& -4.80&  5.79&development\\
GO:0030529&32&4&2.35& -4.53&  5.24&ribonucleoprotein complex\\
GO:0003702&106&14&1.72& -4.43&  5.02&RNA polymerase II transcription factor act.\\
GO:0008565&200&31&0.89& -4.43&  5.03&protein transporter activity\\
GO:0003677&621&198&0.35& -4.39&  4.95&DNA binding activity\\
GO:0007420&16&0&0.00& -4.24&  4.65&brain development\\
GO:0004723&33&0&0.00& -4.17&  4.52&Ca-dep. prot. serine/threonine phosphatase act.\\
GO:0008436&10&0&0.00& -4.15&  4.48&heterogeneous nuclear ribonucleoprotein\\
GO:0004724&32&0&0.00& -4.11&  4.41&Mg-dep. prot. serine/threonine phosphatase act.\\
GO:0008420&32&0&0.00& -4.11&  4.41&CTD phosphatase activity\\
GO:0015071&32&0&0.00& -4.11&  4.41&protein phosphatase type 2C activity\\
GO:0017018&32&0&0.00& -4.11&  4.41&myosin phosphatase activity\\
GO:0030357&31&0&0.00& -4.08&  4.34&protein phosphatase type 2B activity\\
GO:0004674&261&50&1.04& -4.02&  4.24&protein serine/threonine kinase activity\\
GO:0004840&42&0&0.00& -4.01&  4.21&ubiquitin conjugating enzyme activity\\
GO:0007048&230&36&0.56& -3.95&  4.11&oncogenesis\\
GO:0005525&190&28&0.83& -3.94&  4.09&GTP binding activity\\
GO:0008151&210&28&0.33& -3.89&  4.00&cell growth and/or maintenance\\
GO:0006470&95&3&0.03& -3.85&  3.94&protein amino acid dephosphorylation\\
GO:0000158&32&0&0.00& -3.83&  3.89&protein phosphatase type 2A activity\\
GO:0003713&104&8&0.39& -3.82&  3.87&transcription co-activator activity\\
GO:0008380&8&0&0.00& -3.82&  3.88&RNA splicing\\
GO:0016563&32&2&0.83& -3.79&  3.83&transcriptional activator activity\\
GO:0006511&80&4&0.22& -3.76&  3.77&ubiquitin-dependent protein catabolism\\
GO:0003729&8&0&0.00& -3.63&  3.55&mRNA binding activity\\
GO:0045182&193&30&1.01& -3.63&  3.55&translation regulator activity\\
GO:0006886&191&26&0.56& -3.61&  3.52&intracellular protein transport\\
GO:0007399&176&26&1.03& -3.59&  3.47&neurogenesis\\
GO:0006512&56&1&0.06& -3.56&  3.42&ubiquitin cycle\\
GO:0006446&27&1&0.42& -3.53&  3.38&regulation of translational initiation\\
GO:0005737&450&105&0.14& -3.48&  3.30&cytoplasm\\
GO:0006357&135&14&0.55& -3.32&  3.05&regulation of transcription from Pol II promoter\\
\end{tabular}


\begin{thebibliography}{99}
  
\bibitem{MolBioCell} Alberts B, Bray D, Lewis J, Raff, M, Roberts K,
  et al. (1994) Molecular Biology of the Cell.  New York: Garland
  Publishing.
  
\bibitem{Altschul} Altschul SF, Gish W, Miller W, Meyers EW, Lipman DJ
  (1990) Basic Local Alignment Search Tool.  J Mol Biol 215:403.
  
\bibitem{Bernardi} Bernardi G (2000) Isochores and the evolutionary
  genomics of vertebrates.  Gene 241:3.
  
\bibitem{Casane} Casane D, Boissinot S, Chang BH-J, Shimmin LC, Li WH
  (1997) Mutation pattern among regions of the primate genome.  J Mol
  Evol 45:216.
  
\bibitem{Castresana} Castresana J (2002) Genes on human Chromosome 19
  show extreme divergence from the mouse orthologs and a high GC
  content.  Nucl Acids Res 30:1751.
  
\bibitem{DurbinBook} Durbin R, Eddy S, Krogh A, Mitchison G (1998)
  Biological Sequence Analysis. Cambridge:Cambridge University Press.
  
\bibitem{Hovergen} Duret L, Mouchiroud D, Gouy M (1994) HOVERGEN, a
  database of homologous vertebrate genes. Nucleic Acids Res 22:2360.
  
\bibitem{DurettRNA} Duret L, Mouchiroud D (2000) Determinants of
  substitution rates in mammalian genes: expression pattern affects
  selection intensity but not mutation rate.  Mol Bio Evol 17:68.
  
\bibitem{Francino} Francino MP, Ochman H (1997) Strand asymmetries in
  DNA evolution.  Trends in Genetics 13:240.
  
\bibitem{GO} Gene Ontology Consortium (2000) Gene Ontology: tool for
  the unification of biology.  Nature Genet. 25: 25-29.
  
\bibitem{GoldmanYang} Goldman N, Yang Z (1994) A codon-based model of
  nucleotide substitution for protein-coding DNA sequences.  Mol Bio
  Evol 11:725.
  
\bibitem{Kuby} Goldsby R, Kindt T, Osborne B, Kuby J (2000)
  Immunology. New York: W. H. Freeman and Co.
  
\bibitem{GraurLi} Graur D, Li WH (2000) Fundamentals of Molecular
  Evolution, 2nd edition. Sunderland:Sinauer Associates.
  
\bibitem{GuLi} Gu X and Li WH (1994) A model for the correlation of
  mutation rate with GC content and the origin of GC-rich isochores.
  J Mol Evol 38:468.
  
\bibitem{Haussler} Hardison RC, Roskin KM, Yang S, Diekhans M, Kent
  WJ, et al. (2003) Covariation in frequencies of substitution,
  deletion, transposition, and recombination during eutherian
  evolution.  Genome Research 13:13.
  
\bibitem{Hsiao} Hsiao LL, Dangond F, Yoshida T, Hong R, Jensen RV, et
  al.  (2001) A compendium of gene expression in normal human tissues.
  Physiol Genomics 7:97.
  
\bibitem{KHuang} Huang K (1987) Statistical Mechanics. New York: John
  Wiley and Sons.
  
\bibitem{HumanGenome} International Human Genome Sequencing Consortium
  (2001) Initial sequencing and analysis of the human genome.  Nature
  409:860.
  
\bibitem{Iida} Iida K, Akashi H (2000) A test of translational
  selection at `silent' sites in the human genome: base composition
  comparisons in alternatively spliced genes.  Gene 261:93.
  
\bibitem{Kanaya} Kanaya S, Yamada Y, Kinouchi M, Kudo Y, Ikemura T
  (2001) Codon usage and tRNA genes in eukaryotes: correlation of
  codon usage diversity with translation efficiency and with
  CG-dinucleotide usage as assessed by multivariate analysis.  J Mol
  Evol 53:290.
  
\bibitem{Kumar} Kumar S, Subramanian S (2002) Mutation rates in
  mammalian genomes.  Proc Natl Acad Sci 99:803.
  
\bibitem{Lane} Lane RP, Cutforth T, Young J, Athanasiou M, Friedman C,
  et al. (2001) Genomic analysis of orthologous mouse and human
  olfactory receptor loci.  Proc Natl Acad Sci 98:7390.
  
\bibitem{LercherWilliams} Lercher MJ, Williams EJB, and Hurst LD
  (2001) Local similarity in evolutionary rates extends over whole
  chromosomes in human-rodent and mouse-rate comparisons: implications
  for understanding the mechanistic basis of the male mutation bias.
  Mol Biol Evol 18:2032.
  
\bibitem{Lercher} Lercher MJ, Hurst LD (2002) Human SNP variability
  and mutation rate are higher in regions of high recombination.
  Trends in Genetics 18:337.
  
\bibitem{WHLi} Li WH (1993) Unbiased estimation of the rates of
  synonymous and nonsynonymous substitution.  J Mol Evol 36:96.
  
\bibitem{Matassi} Matassi G, Sharp PM, Gautier C (1999) Chromosomal
  location effects on gene sequence evolution in mammals.  Current
  Biology 9:786.
  
\bibitem{Mouchiroud} Mouchiroud D, Gautier C, Bernardi G (1995)
  Frequences of Synonymous Substitutions in Mammals Are Gene-Specific
  and Correlated with Frequencies of Nonsynonymous Substitutions. J
  Mol Evol 40:107.
  
\bibitem{MGSConsortium} Mouse Genome Sequencing Consortium (2002)
  Initial sequencing and comparative analysis of the mouse genome.
  Nature 420:520.
  
\bibitem{NeiGojobori} Nei M, Gojobori T (1986). Simple methods for
  estimating the numbers of synonymous and nonsynonymous nucleotide
  substitutions.  Mol Bio Evol 3:418-426.
  
\bibitem{Ohta} Ohta T, Ina Y (1995). Variation in synonymous
  substitution rates among mammalian genes and the correlation between
  synonymous and nonsynonymous divergences. J Mol Evol 41:717.
  
\bibitem{Pal} Pal C, Hurst LD (2003). Evidence for co-evolution of
  gene order and recombination rate. Nature Genetics 33:392.
  
\bibitem{Papavasiliou} Papavasiliou FN, Schatz DG (2002) Somatic
  hypermutation of immunoglobulin genes: merging mechanisms for
  genetic diversity.  Cell 109:S35.
  
\bibitem{Percudani} Percudani R, Ottonello S (1999) Selection at the
  wobble position of codons read by the same tRNA in Saccharomyces
  cerevisiae.  Mol Bio Evol 16:1752.
  
\bibitem{Perry} Perry J, Ashworth A (1999) Evolutionary rate of a gene
  affected by chromosomal position.  Current Biology 9:987.
  
\bibitem{Rice} Rice J (1995) Mathematical Statistics and Data
  Analysis.  Belmont: Duxbury Press.
  
\bibitem{Rouquier} Rouquier S, Taviaux S, Trask BJ, Brand-Arpon V, van
  den Engh G, et al. (1998) Distribution of olfactory receptor genes
  in the human genome.  Nature Genetics 18:243.
  
\bibitem{Sharon} Sharon D, Glusman G, Pilpel Y, Khen M, Gruetzner F,
  et al. (1999) Primate evolution of an olfactory receptor cluster:
  diversification by gene conversion and recent emergence of
  pseudogenes. Genomics 61:24.
  
\bibitem{Sharp} Sharp PM, Averof M, Lloyd AT, Matassi G, and Peden JF
  (1995) DNA sequence evolution: the sounds of silence. Phil Trans R
  Soc Lond B 349:241.
  
\bibitem{SharpLi} Sharp PM, Li WH (1987) The rate of synonymous
  substitution in enterobacterial genes is inversely related to codon
  usage bias.  Mol Bio Evol 4:222.
  
\bibitem{SmithHurst} Smith NGC, Hurst LD (1999a) The causes of
  synonymous rate variation in the rodent genome: can substitution
  rates be used to estimate the sex bias in mutation rate?  Genetics
  152:661.
  
\bibitem{SmithHurstb} Smith NGC, Hurst LD (1999b) The effect of tandem
  substitutions on the correlation between synonymous and
  nonsynonymous rates in rodents. Genetics 153:1395.
  
\bibitem{TamuraNei} Tamura K, Nei M (1993) Estimation of the number of
  nucleotide substitutions in the control region of mitochondrial DNA
  in humans and chimpanzees.  Mol Bio Evol 10:512.
  
\bibitem{Tasic} Tasic B, Nabholz CE, Baldwin KK, Kim Y, Rueckert EH,
  et al. (2002) Promoter choice determines splice site selection in
  protocadherin alpha and gamma pre-mRNA splicing.  Mol Cell 10:21.
  
\bibitem{REV} Tavere S (1986) Some probabilistic and statistical
  problems on the analysis of DNA sequences.  Lec Math Life Sci 17:57.
  
\bibitem{Uemura} Uemura T (1998) The cadherin superfamily at the
  synapse: more members, more missions.  Cell 93:1095.
  
\bibitem{WilliamsHurstb} Williams EJB and Hurst LD (2002) Clustering
  of tissue-specific genes underlies much of the similarity in rates
  of protein evolution of linked genes. J Mol Evol 54:511.
  
\bibitem{Winzeler} Winzeler EA, Shoemaker DD, Astromoff A, Liang H,
  Anderson K, et al. (1999) Functional characterization of the
  Saccharomyces cerevisiae genome by gene deletion and parallel
  analysis.  Science 285:901.
  
\bibitem{Wolfe} Wolfe KH, Sharp PM, Li WH (1989) Mutation rates differ
  among regions of the mammalian genome.  Nature 337:283.
  
\bibitem{QWu99} Wu Q, Zhang T, Cheng J-F, Kim Y, Grimwood J, et al.
  (2001) Comparative DNA sequence analysis of mouse and human
  protocadherin gene clusters.  Genome Research 11:389.
  
\bibitem{Yang} Yang Z (1997) PAML: a program package for phylogenetic
  analysis by maximum likelihood. Comput Appl Biosci 13:555.
  
\bibitem{Zhang} Zhang L and Li WH (2003) Mammalian housekeeping genes
  evolve more slowly than tissue-specific genes. Mol Bio Evol
  epub:http://mbe.oupjournals.org/cgi/reprint/msh010v1.


\end{thebibliography}
\end{document}